\documentclass[sigconf]{acmart}

\AtBeginDocument{%
  }

\setcopyright{none}

\acmConference[FAccTRec@RecSys ’24]{7th FAccTRec Workshop on Responsible Recommendation at RecSys 2024}{October 14--18,
  2024}{Bari, Italy}

\begin{document}

\title{Bypassing the Popularity Bias: Repurposing Models for Better Long-Tail Recommendation}

\author{Václav Blahut}
\email{vaclav.blahut@firma.seznam.cz}
\affiliation{%
  \institution{Seznam.cz, a.s.}
  \city{Prague}
  \country{Czech Republic}
}

\author{Karel Koupil}
\email{karel.koupil@firma.seznam.cz}
\affiliation{%
  \institution{Seznam.cz, a.s.}
  \city{Prague}
  \country{Czech Republic}
}


\begin{abstract}
Recommender systems play a crucial role in shaping information we encounter online, whether on social media or when using content platforms, thereby influencing our beliefs, choices, and behaviours. Many recent works address the issue of fairness in recommender systems, typically focusing on topics like ensuring equal access to information and opportunities for all individual users or user groups, promoting diverse content to avoid filter bubbles and echo chambers, enhancing transparency and explainability, and adhering to ethical and sustainable practices. 

In this work, we aim to achieve a more equitable distribution of exposure among publishers on an online content platform, with a particular focus on those who produce high quality, long-tail content that may be unfairly disadvantaged. We propose a novel approach of repurposing existing components of an industrial recommender system to deliver valuable exposure to underrepresented publishers while maintaining high recommendation quality. To demonstrate the efficiency of our proposal, we conduct large-scale online AB experiments, report results indicating desired outcomes and share several insights from long-term application of the approach in the production setting.
\end{abstract}




\maketitle

\section{Introduction}
\label{sec:intro}

\begin{figure}[h]
  \centering
  \includegraphics[width=\linewidth]{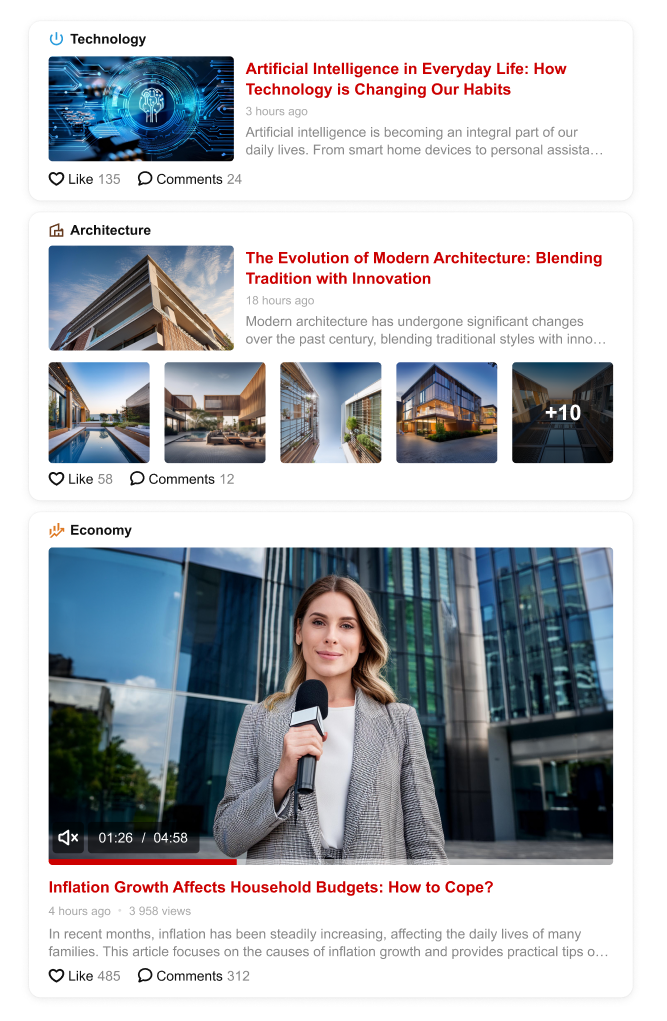}
  \caption{Illustration of our endless feed of content.}
  \Description{Illustration of our endless feed of content.}
  \label{fig:feed}
\end{figure}

Our online content platform serves millions of daily active users and recommends content produced by many different publishers. Users are offered a personalized selection of online media content, consisting of news and entertainment articles, videos and podcasts, as illustrated in a Figure~\ref{fig:feed}. The platform is privately owned and the majority of its revenue comes from serving online advertisements.

As the most visited content platform in our country, we make great demands on all publishers in terms of quality of the content itself and impose rules for displaying an advertisement. Long-tail content \footnote{The term \textit{long-tail content}~\cite{yin2012challenginglongtailrecommendation} refers to the items that are less popular compared to the most popular items. It depends on the shape of exposure distribution curve and often constitutes the majority of items, in contrast to the small number of very popular items.} with high specificity and quality, designed for a smaller audience, is often expensive to produce and in our recommender system suffers from a popularity bias that leads to less attention due to fewer clicks and a lower click-through rate (CTR). As a result, more specific and less popular publishers may not thrive, and being involved in our platform may not be economically beneficial for them in the long-term. This can be regarded as a systemic bias in our setting.  

Recommending long-tail items is generally considered to be valuable to the users \cite{anderson_2006}, as these are items that users are less likely to know about. Additionally, exposing users to these items can increase diversity and serendipity of a recommendation list, which surprises and satisfies the users because properly selected long-tail items can still be very relevant to them \cite{5680904, 10.1145/3123266.3123316}.

A market that suffers from a popularity bias will lack opportunities to discover more obscure products and will be, by definition, dominated by a few large brands or well-known artists \cite{10.1145/1722149.1722154}. In our work, we try to mitigate this bias as well as strive to improve the recommendation of the long-tail content to users.

A typical recommender system searches for the best items for each user in terms of a defined objective.  We decided to reverse this approach and formalize the problem as follows: \textit{for each item of the selected publisher, we search for its best users}.  We believe and further prove that this approach can mitigate the aforementioned systemic bias and provides sufficient revenue for selected long-tail publishers.

Since our recommender system recommends an item to a user and we want to promote a long-tail group of publishers, we need to create a mechanism that aligns our approach to the publisher level.  This involves setting a minimum exposure, which is the amount of exposure each of the items produced by selected publishers should receive, that is considered sufficient and is constant across all items. In our experiments, exposure of an item is represented by the number of visible impressions, that is, the number of times a given item has been recommended to a user and visibly displayed on user’s screen for defined period of time. To establish the value of minimum exposure, we take into account typical revenue of an advertisement that is displayed on a selected publisher’s website, as well as average CTR of an item.

In summary, we make the following contributions: 

\begin{itemize}
    \item we propose a novel approach to recommending long-tail items,
    \item we employ several system-level fairness metrics,
    \item we conduct multiple online experiments, including an ablation study
    \item we investigate the effectiveness of the proposed approach in terms of fairness and performance metrics
    \item we incorporate the proposed method into our standard recommender system and report long-term experience 
\end{itemize}

\section{Related work}

In order to measure the effect of the proposed approach on equitable resource distribution at our platform, system-level fairness metrics need to be defined. Lazovich et al. \cite{Lazovich_2022} present the distributional inequality metrics, a set of metrics originating from economics, and define desirable criteria to evaluate their ability to measure disparities in content exposure. Diaz et al. \cite{Diaz_2020} define a set of metrics based on the principle of equal expected exposure, which takes into account the relevance of an item.

A common way of improving system-level fairness in recommender systems is via randomization. Diaz et al. \cite{Diaz_2020} use Plackett-Luce sampling, which samples permutations based on the scores from the retrieval or the ranking model. Another method, called Rank Transposition, ignores the scores and randomly shuffles the original ranked list. Bower et al. \cite{bower2022randomisntfaircandidate} show that, in industrial settings, randomization only at one specific stage of recommendation process may not lead to improvements in fairness, and that it is necessary to treat fairness holistically, at all stages of possibly very complex recommender system. To do so, authors present Plackett-Luce sampling with Inverse Candidate Frequency Weights, which helps mitigate possible biases arising from retrieval stage of recommendation. 

Maximum Inner Product Search (k-MIPS) is a well-studied problem from the area of collaborative filtering recommendation \cite{rendle2020neuralcollaborativefilteringvs}. The idea of reversing the k-MIPS to efficiently find suitable users for an item was first proposed by Amagata and Hara \cite{amagata2021reversemaximuminnerproduct}. The authors define the reverse k-MIPS as the problem of finding a set of k users with maximum inner product for a query item, explore existing exact and approximate search algorithms and propose their own search algorithm called Simpfer, which they further improve in \cite{10.1145/3587215}.

\section{Inverse Retrieval Model}

Given an item, our goal is to retrieve an ordered set of users who are likely to find the item relevant and to consume it. To achieve that, we suggest repurposing the two-tower retrieval model \cite{50257} that is already trained and used in our main recommendation pipeline. In a standard setting of recommending items for a single user, the model is used to generate user and item embeddings. Single user embedding is then used as a query in an Approximate Nearest Neighbors (ANN) index of item embeddings to retrieve the top N most similar items. To tackle our problem of retrieving users for an item, we propose to invert this setting. Given a set of users, we use the model to generate an embedding for each of the users and store those embeddings in the ANN index. Then, we use the embedding of the given item to query this index and retrieve the top N most similar users. Note that any other model optimized for user-item embedding (dis-)similarity can be used as a replacement. 

\begin{figure}[b]
  \centering
  \includegraphics[width=\linewidth]{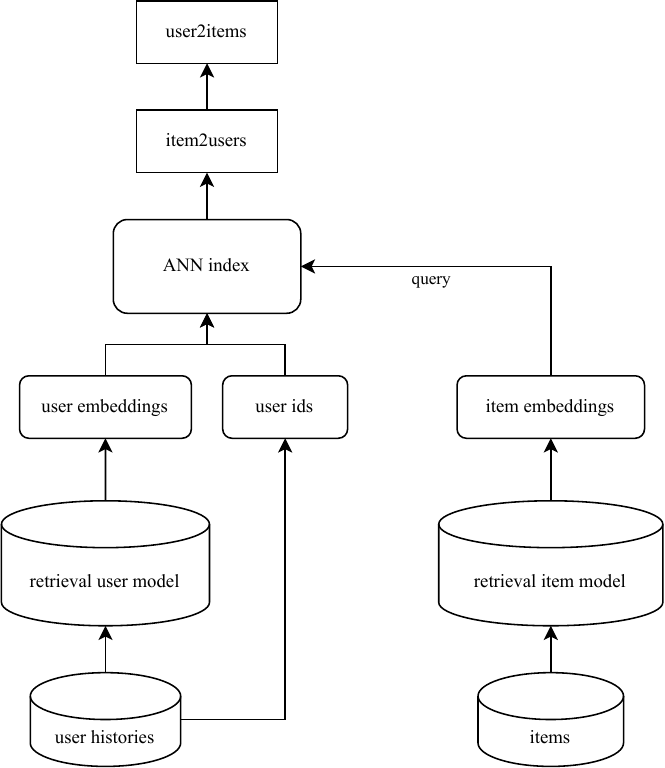}
  \caption{InvR pipeline diagram in production conditions}
  \Description{InvR pipeline diagram in production conditions}
  \label{fig:ir_scheme}
\end{figure}

This Inverse Retrieval (InvR) process can be repeated for all selected long-tail items that did not receive enough exposure at given time. The detailed diagram of the pipeline is shown in Figure~\ref{fig:ir_scheme}. By its nature, this approach fits better in offline recommendation setting, so all item-to-users candidates are periodically recomputed, transposed to user-to-items (single user can appear as a candidate for multiple items) and presented to the user once they use the service. Note that the number of users per item is a crucial hyperparameter that has to be carefully selected according to the specific context. The variables that affect the selection include but are not limited to the required minimum item exposure, the number of items and users, the frequency of candidate recomputation and the lifetime of an item, if any kind of item expiration is present. 

\subsection{Ordering items for single user}
\label{sec:ordering}

In cases when there are multiple items recommended for a single user, we need to decide the ordering of items presented to the user. The most apparent solution is to use the score provided by the ANN index, in our case dot product of the user and the item embedding. This approach, however, can be strongly biased by item popularity as more popular items often gain higher scores \cite{10.1145/3523227.3546757} regardless of the actual relevance to the user, defeating the very purpose of recommending long-tail items. A more fairness-oriented solution is to randomly shuffle items. We argue that the most appropriate method is to order the items by the rank of a given user with respect to a given item. In other words, if user ranks high for the given item, compared to all other users according to the score, then this item should be presented to this user at a high position, ignoring the absolute value of the score. This way, the priority is to present the given item to the most promising users first. 

\subsection{Considerations and limitations}

One may argue that an equivalent of item popularity bias may exist in the set of users, heavy users being ranked high in similar manner as popular items are in standard recommender system scenarios. Such problem may occur if the users were represented by their unique ID. In our system, it is naturally avoided by representing the users by their recent interaction history, truncated to a fixed length, so the rank of a user should not depend on their level of activity. Also, to limit the number of item candidates per user during the user-to-item transposition, we retrieve more user candidates for each item, sort all user-item candidate pairs by score or rank (depending on the variant) and in a single pass, we assign items to users until predefined item per user limit is reached, after which the user is skipped for further candidates. This way, the items are distributed among more users instead of accumulating on fewer heavy users and every item is guaranteed to be assigned exactly to the required number of users. 

Another challenge is the quality of cold start item embeddings. Since the ID-based item embeddings in the two-tower retrieval model are trained on user-item interactions, the representations of new items with no interactions are virtually random and useless for finding relevant users. We employ standard randomization solution to the item cold start by randomly inserting new items into recommendation slates for a pre-defined period of time or until sufficient number of interactions have been gathered.

\section{Setup}

\subsection{Recommender system description}
\label{sec:rec_sys_desc}

Our main recommendation pipeline follows the 4-stage framework as described by Higley et al. \cite{10.1145/3523227.3551468}. The retrieval and ranking stages utilize models from TensorFlow Recommenders library\footnote{https://www.tensorflow.org/recommenders/}, specifically the two-tower model for retrieval and Deep and Cross Network V2\cite{Wang_2021} for the ranking stage. Both retrieval and ranking models are trained incrementally every 5 minutes. An endless feed of recommended items consists of 20-item slates, incrementally generated online as user scrolls through the feed. The system serves millions of daily active users, handling thousands of requests per second with the latency limits in order of lower hundreds of milliseconds.

The retrieval model, which is also repurposed for InvR, uses hashed ID-based item embeddings. An embedding of user is determined as an average pooling of item embeddings present in given user’s history. Clicked items are used as positive examples and visible but not clicked items as negative. The dimensionality of embeddings is 128, the batch size is 128. We train the model for 10 epochs with learning rate 0.1 and AdaGrad optimizer. All retrieval model hyperparameters were optimized previously during series of experiments and have not been further optimized for the InvR usage. Both item index for retrieval and user index for InvR use ScANN\cite{avq_2020} library. 

\subsection{Incorporating InvR items in recommendation}

The offline InvR pipeline is executed every 60 minutes. During the online recommendation, the pre-generated InvR candidate items are inserted into the slate after the ranking stage and most of the business logic, so the recommended items are isolated from all possible biases arising from the main recommender, as described in Section~\ref{sec:intro}. We argue that the main bias our approach bypasses is the popularity bias. The only remaining business logic applied to InvR-recommended items is simple deduplication. We dedicate up to 3 positions in a specified range of each 20-item slate for InvR recommendation. If an item has been visibly displayed to a user twice, we exclude it for all future visits of the same user. Once an item reaches the minimum exposure, it is no longer supported by the InvR mechanism, though it may still be organically recommended by the main recommender pipeline.

\subsection{Publisher selection}

We select those publishers that suffer from low attention/revenues due to the popularity bias as described in Section~\ref{sec:intro}. For this purpose, we design several business and performance criteria that have to be met in a 6-months window:  low total revenue per month, low average number of visible impressions across publisher's items, a low number of clicks in total per month and finally low revenue per each item on average. Besides previous conditions, publishers have to belong to a group of “original niche” which is characterized by a high level of content production quality and is specialized utterly in one content topic.

\subsection{User selection}

We want to show long-tail content recommendations to the users with stable identity, i.e. those who have sufficiently long history and who consent with personalization. Since the success of our proposal heavily depends on the user actually visiting the platform in the near future so that the InvR-generated item candidates can be presented to them, we select only the users who visited the platform frequently enough in recent past. 

\subsection{A/B testing}

In order to assess the performance of our recommender system, we perform randomized A/B tests on live traffic. For each of the tested variants, we include A/A variant. The size of each variant as well as test duration are chosen empirically w.r.t. statistical significance. In the context of this work, it would be interesting to perform AB testing in terms of items. However, due to the production environment, the unit of randomization in our AB testing system is the user.

\section{Evaluation and discussion}

\begin{figure}
  \includegraphics[width=\linewidth]{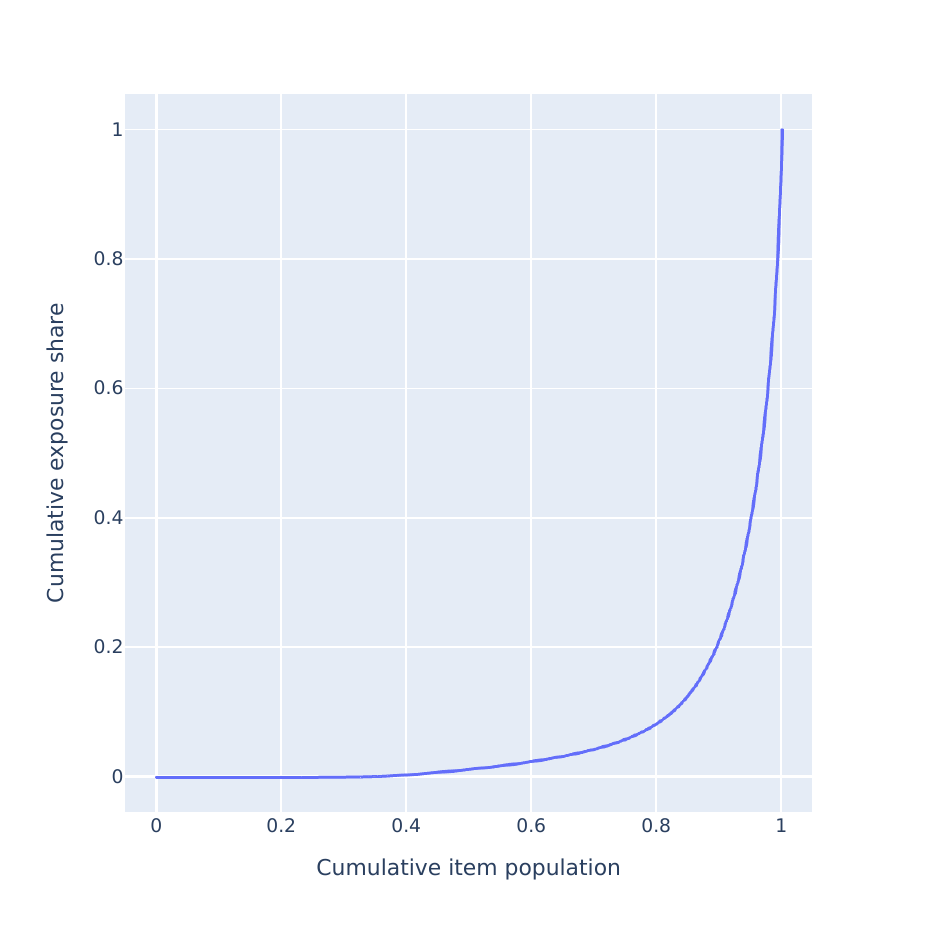}
  \caption{Real-world example of Lorenz curve.}
  \Description{Real-world example of Lorenz curve.}
  \label{fig:lorenz_curve}
\end{figure}

\subsection{Metrics}
\label{sec:metrics}

To evaluate the impact of the proposed method on the exposure of long-tail items, we employ these system-level item-wise fairness metrics: 

a) \textbf{bottom 50 percent share (B50PS)} is a metric derived from Lorenz curve, inspired by top X percent share metric presented by Lazovich et al.\cite{Lazovich_2022}, modified to capture changes in the long-tail items region of the Lorenz curve. It shows the portion of exposure received by the least popular half of the whole item set. While the choice of 50th percentile is arbitrary and depends on the skewness of the Lorenz curve, it is a strict lower-bound of what can be considered a long-tail of a distribution, since even in the edge case of uniform distribution, it is equal to the percentage of equal share. Therefore, in many real-world scenarios such as can be seen in Figure~\ref{fig:lorenz_curve}, where the Lorenz curve is heavily skewed, the bottom half of the item set can safely be considered a long-tail. 

b) \textbf{percentage of sufficiently exposed items (PSEI)} is the percentage of the treated items that have received more than the pre-defined minimum exposure. 

The two selected metrics above are meant to be complementary – B50PS focuses on all items while PSEI captures changes only for InvR-treated items.

The overall amount of users and their attention is beyond our control, making the overall exposure relatively fixed. Consequently, promoting a subset of items will inevitably come at the expense of the rest of the items, especially the most popular ones. To illustrate the impact on the most popular items exposition, we report the aforementioned \textbf{top 1 percent share (T1PS)}\cite{Lazovich_2022}.

To compare the InvR variants described in Section~\ref{sec:variants} with each other and with the ablation study baseline in terms of relevance, we report average \textbf{click-through rate (CTR)} and average number of \textbf{clicks}, both per user, filtered to InvR-generated recommendations only. All reported metric changes are in relative scale. 

\begin{figure*}[t]
  \includegraphics[width=\textwidth]{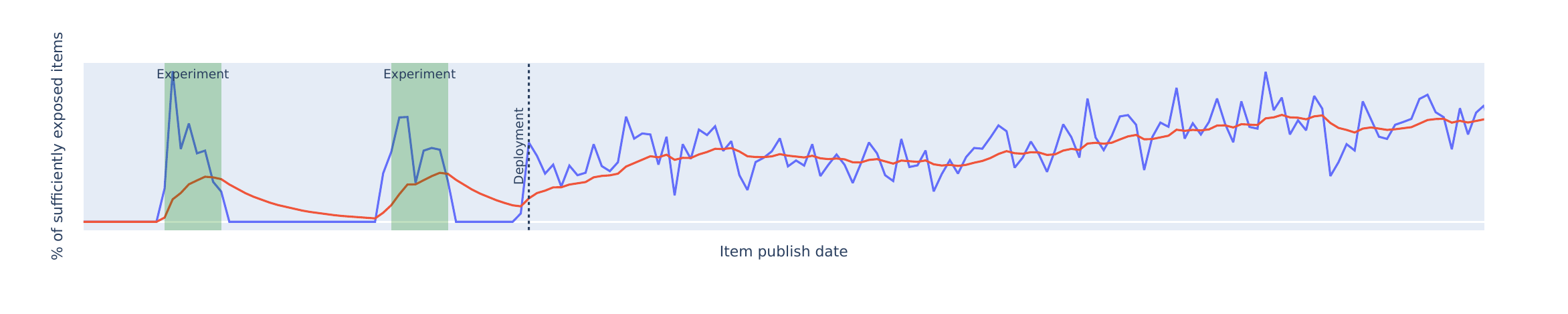}
  \caption{Percentage of sufficiently exposed items (PSEI) over time.}
  \Description{Percentage of sufficiently exposed items (PSEI) over time.}
  \label{fig:psei}
\end{figure*}

\subsection{Variants}
\label{sec:variants}

The \textbf{Baseline} is our recommender system as described in Section~\ref{sec:rec_sys_desc}, without any usage of the InvR. We conduct an ablation study to investigate the performance gain of using the model in InvR pipeline by including \textbf{Random} variant that acts as a baseline for InvR. In this variant, the same conditions apply (set of items and users, periodical offline batch computation and incorporation into the slate), but instead of using the model for user selection, we pick users for an item randomly. Finally, we use three InvR variants described in Section~\ref{sec:ordering} to show differences in the item orderings. \textbf{InvR Random} orders the item candidates for a user randomly, \textbf{InvR Score} orders the candidates by the absolute value of the score and \textbf{InvR User rank} by the user rank w.r.t. the item. 

\begin{table}[t]
  \caption{Experimental results}
  \label{tab:results}
  \resizebox{\columnwidth}{!}{
    \begin{tabular}{lccccc}
        \toprule
        Variant & B50PS & PSEI & T1PS & CTR (InvR) & Clicks (InvR) \\
        \midrule
        Baseline & 0 \% & 0 \% & 0 \% & - & - \\
        Random & +9.2 \% & +181 \% & -1.0 \% & 0 \% & 0 \% \\
        InvR Random & +9 \% & +43 \% & -0.5 \% & +271 \% & +106 \% \\
        InvR Score & +9.9 \% & +41 \% & -0.9 \% & +261 \% & +103 \% \\
        InvR User rank & +33.3 \% & +45 \% & -1.6 \% & +300 \% & +120 \% \\
        \bottomrule
    \end{tabular}
    }
\end{table}

\subsection{Overall results}

See Table~\ref{tab:results} for the experimental result. The metric bottom 50 percent share, B50PS shows that while all variants have boosted the exposure of long-tail items, the InvR User rank variant surpassed others by a large margin. This metric is computed on all items, not only those produced by selected publishers. As for the percentage of sufficiently exposed items metric, PSEI, where only items from selected publishers are considered, we can see that the Random variant was outstandingly successful. This comes as no surprise, because - in the Random variant - items are uniformly distributed between all users, leading to maximal exposure, while all other InvR methods focus on a much smaller set of relevant users. The cost of the high exposure is a severely low CTR and Clicks when compared to other InvR variants, as a long-tail item delivered to a random user will most likely be irrelevant to the user. The discrepancies between B50PS and PSEI for Random and InvR variants are consequent upon the fact that each metric is measured over a different set of items (see the mention of the complementarity of these two metrics in Section~\ref{sec:metrics}).

In terms of the impact on the most popular items exposition captured by top 1 percent share, T1PS, where a reduction is (from a fairness point of view) considered positive change, the InvR User rank variant significantly outperformed others.

Focusing on the CTR and Clicks, we found it surprising that the InvR Score variant did not surpass the others thanks to the popularity bias in scores, which remains unexplained. We conclude that the InvR User rank item ordering method is consistently the right approach.

Since InvR effectively pushes down other, possibly more popular and clickable items, it is natural that relevance-focused KPIs may suffer. To illustrate the cost of deploying the winning InvR User rank variant to production, we report 1.04 \% drop in overall user average CTR and 1.67 \% decrease in overall average number of clicks per user. 

\subsection{Further experiments}

After the main experiment where described variants were tested, a few more experiments followed. For example, we examined the impact of position bias on item exposure by modifying the range of positions where InvR items are inserted into slate, confirming that by placing item on higher position, the exposure increases significantly, with significantly negative impact on KPIs. We also experimented with different user counts per item, in which case we discovered that the increase in exposure is sublinear to the increase of users per item. Other experiments tested various user-item allocation algorithms that assign candidates more evenly in edge cases (leading to increase in exposure of the least popular items), inclusion of unregistered users (no significant impact) and loosening the criteria for publisher selection (lower overall exposure per item). 

\subsection{Mentionable side-effects and long-term experience}

Apart from the results above, other interesting details surfaced when we analysed the experiment. The main recommender pipeline secondarily benefited in terms of diversity (measured by total number of unique recommended items) thanks to the fact that the InvR-generated data was used for training the main model. The number of unique items recommended by the main pipeline in InvR variants increased by 18 \% when compared to Baseline.

Finally, the InvR User rank variant was deployed as our new production baseline. Over the course of the following months, we observed interesting behaviors in the system. While there was an immediate increase in PSEI metric after the deployment of InvR, the PSEI continued to grow until it reached similar values as the Random variant did in our main experiment, while maintaining good performance in terms of KPIs. The increase of global PSEI after deployment, as well as during the experiments, can be seen in Figure~\ref{fig:psei}. The line in blue is the actual daily value and the red line is exponentially weighted moving average with $\alpha=0.125$.

We also noticed that on long-term average, only about 40 \% of minimum exposure is generated by InvR itself until it gets shut off for given item due to reaching the minimum exposure, suggesting that InvR helps ``kick-start'' the item and the main recommender takes care of the rest.

\section{Conclusion and future work}

In this paper, we proposed a novel way of repurposing part of industrial recommender system to tackle the problem of long-tail content recommendation, which we argue is mainly caused by popularity bias. After describing the method and all its variants, we reported the results of online experiments, showing significant increase in exposure of selected long-tail publishers, accompanied by impacts on the KPIs. We discussed some further observations and side-effects as well as long-terms experience after successful deployment to production. 

In conclusion, proposed approach meets the needs of two primary stakeholders: it supplies publishers with sufficient and relevant traffic and delivers well-personalized recommendations to users, resulting in a positive long-term impact on the platform stability from business perspective.  

There are multiple directions of possible future work. The InvR may greatly benefit from a more sophisticated solution of item cold start, applicable to the two-tower retrieval model, which would be to replace item ID-based embedding with pre-trained content-based one, further trained on user-item interactions. We may also perform the model hyperparameter optimization for the InvR use case or experiment with completely different user and item embedding models. To rule out the influence of selection of long-tail publishers, we may conduct an experiment where all items receive the InvR treatment. We would also like to focus more on the impact of our solution on the user experience and user-centred fairness. 

\begin{acks}
We thank Radek Tomšů, Vít Líbal, Milan Vancl, Jaroslav Kuchař and Jan Vršovský for their extensive feedback on this paper. We thank Josef Florian for providing important details of the production system.
\end{acks}

\bibliographystyle{ACM-Reference-Format}
\bibliography{base}


\begin{thebibliography}{16}


\ifx \showCODEN    \undefined \def \showCODEN     #1{\unskip}     \fi
\ifx \showDOI      \undefined \def \showDOI       #1{#1}\fi
\ifx \showISBNx    \undefined \def \showISBNx     #1{\unskip}     \fi
\ifx \showISBNxiii \undefined \def \showISBNxiii  #1{\unskip}     \fi
\ifx \showISSN     \undefined \def \showISSN      #1{\unskip}     \fi
\ifx \showLCCN     \undefined \def \showLCCN      #1{\unskip}     \fi
\ifx \shownote     \undefined \def \shownote      #1{#1}          \fi
\ifx \showarticletitle \undefined \def \showarticletitle #1{#1}   \fi
\ifx \showURL      \undefined \def \showURL       {\relax}        \fi
\providecommand\bibfield[2]{#2}
\providecommand\bibinfo[2]{#2}
\providecommand\natexlab[1]{#1}
\providecommand\showeprint[2][]{arXiv:#2}

\bibitem[Adomavicius and Kwon(2012)]%
        {5680904}
\bibfield{author}{\bibinfo{person}{Gediminas Adomavicius} {and} \bibinfo{person}{YoungOk Kwon}.} \bibinfo{year}{2012}\natexlab{}.
\newblock \showarticletitle{Improving Aggregate Recommendation Diversity Using Ranking-Based Techniques}.
\newblock \bibinfo{journal}{\emph{IEEE Transactions on Knowledge and Data Engineering}} \bibinfo{volume}{24}, \bibinfo{number}{5} (\bibinfo{year}{2012}), \bibinfo{pages}{896--911}.
\newblock
\urldef\tempurl%
\url{https://doi.org/10.1109/TKDE.2011.15}
\showDOI{\tempurl}


\bibitem[Amagata and Hara(2021)]%
        {amagata2021reversemaximuminnerproduct}
\bibfield{author}{\bibinfo{person}{Daichi Amagata} {and} \bibinfo{person}{Takahiro Hara}.} \bibinfo{year}{2021}\natexlab{}.
\newblock \bibinfo{title}{Reverse Maximum Inner Product Search: How to efficiently find users who would like to buy my item?}
\newblock
\newblock
\showeprint[arxiv]{2110.07131}~[cs.DB]
\urldef\tempurl%
\url{https://arxiv.org/abs/2110.07131}
\showURL{%
\tempurl}


\bibitem[Amagata and Hara(2023)]%
        {10.1145/3587215}
\bibfield{author}{\bibinfo{person}{Daichi Amagata} {and} \bibinfo{person}{Takahiro Hara}.} \bibinfo{year}{2023}\natexlab{}.
\newblock \showarticletitle{Reverse Maximum Inner Product Search: Formulation, Algorithms, and Analysis}.
\newblock \bibinfo{journal}{\emph{ACM Trans. Web}} \bibinfo{volume}{17}, \bibinfo{number}{4}, Article \bibinfo{articleno}{26} (\bibinfo{date}{jul} \bibinfo{year}{2023}), \bibinfo{numpages}{23}~pages.
\newblock
\showISSN{1559-1131}
\urldef\tempurl%
\url{https://doi.org/10.1145/3587215}
\showDOI{\tempurl}


\bibitem[Anderson(2006)]%
        {anderson_2006}
\bibfield{author}{\bibinfo{person}{Chris Anderson}.} \bibinfo{year}{2006}\natexlab{}.
\newblock \bibinfo{booktitle}{\emph{{The long tail : why the future of business is selling less of more}}}.
\newblock \bibinfo{publisher}{Hyperion}.
\newblock
\showISBNx{1401302378}


\bibitem[Bower et~al\mbox{.}(2022)]%
        {bower2022randomisntfaircandidate}
\bibfield{author}{\bibinfo{person}{Amanda Bower}, \bibinfo{person}{Kristian Lum}, \bibinfo{person}{Tomo Lazovich}, \bibinfo{person}{Kyra Yee}, {and} \bibinfo{person}{Luca Belli}.} \bibinfo{year}{2022}\natexlab{}.
\newblock \bibinfo{title}{Random Isn't Always Fair: Candidate Set Imbalance and Exposure Inequality in Recommender Systems}.
\newblock
\newblock
\showeprint[arxiv]{2209.05000}~[cs.IR]
\urldef\tempurl%
\url{https://arxiv.org/abs/2209.05000}
\showURL{%
\tempurl}


\bibitem[Celma and Cano(2008)]%
        {10.1145/1722149.1722154}
\bibfield{author}{\bibinfo{person}{\`{O}scar Celma} {and} \bibinfo{person}{Pedro Cano}.} \bibinfo{year}{2008}\natexlab{}.
\newblock \showarticletitle{From hits to niches? or how popular artists can bias music recommendation and discovery}. In \bibinfo{booktitle}{\emph{Proceedings of the 2nd KDD Workshop on Large-Scale Recommender Systems and the Netflix Prize Competition}} (Las Vegas, Nevada) \emph{(\bibinfo{series}{NETFLIX '08})}. \bibinfo{publisher}{Association for Computing Machinery}, \bibinfo{address}{New York, NY, USA}, Article \bibinfo{articleno}{5}, \bibinfo{numpages}{8}~pages.
\newblock
\showISBNx{9781605582658}
\urldef\tempurl%
\url{https://doi.org/10.1145/1722149.1722154}
\showDOI{\tempurl}


\bibitem[Diaz et~al\mbox{.}(2020)]%
        {Diaz_2020}
\bibfield{author}{\bibinfo{person}{Fernando Diaz}, \bibinfo{person}{Bhaskar Mitra}, \bibinfo{person}{Michael~D. Ekstrand}, \bibinfo{person}{Asia~J. Biega}, {and} \bibinfo{person}{Ben Carterette}.} \bibinfo{year}{2020}\natexlab{}.
\newblock \showarticletitle{Evaluating Stochastic Rankings with Expected Exposure}. In \bibinfo{booktitle}{\emph{Proceedings of the 29th ACM International Conference on Information \& Knowledge Management}} (Virtual Event, Ireland) \emph{(\bibinfo{series}{CIKM '20})}. \bibinfo{publisher}{Association for Computing Machinery}, \bibinfo{address}{New York, NY, USA}, \bibinfo{pages}{275–284}.
\newblock
\showISBNx{9781450368599}
\urldef\tempurl%
\url{https://doi.org/10.1145/3340531.3411962}
\showDOI{\tempurl}


\bibitem[Guo et~al\mbox{.}(2020)]%
        {avq_2020}
\bibfield{author}{\bibinfo{person}{Ruiqi Guo}, \bibinfo{person}{Philip Sun}, \bibinfo{person}{Erik Lindgren}, \bibinfo{person}{Quan Geng}, \bibinfo{person}{David Simcha}, \bibinfo{person}{Felix Chern}, {and} \bibinfo{person}{Sanjiv Kumar}.} \bibinfo{year}{2020}\natexlab{}.
\newblock \showarticletitle{Accelerating Large-Scale Inference with Anisotropic Vector Quantization}. In \bibinfo{booktitle}{\emph{International Conference on Machine Learning}}.
\newblock
\urldef\tempurl%
\url{https://arxiv.org/abs/1908.10396}
\showURL{%
\tempurl}


\bibitem[Higley et~al\mbox{.}(2022)]%
        {10.1145/3523227.3551468}
\bibfield{author}{\bibinfo{person}{Karl Higley}, \bibinfo{person}{Even Oldridge}, \bibinfo{person}{Ronay Ak}, \bibinfo{person}{Sara Rabhi}, {and} \bibinfo{person}{Gabriel de Souza Pereira~Moreira}.} \bibinfo{year}{2022}\natexlab{}.
\newblock \showarticletitle{Building and Deploying a Multi-Stage Recommender System with Merlin}. In \bibinfo{booktitle}{\emph{Proceedings of the 16th ACM Conference on Recommender Systems}} (Seattle, WA, USA) \emph{(\bibinfo{series}{RecSys '22})}. \bibinfo{publisher}{Association for Computing Machinery}, \bibinfo{address}{New York, NY, USA}, \bibinfo{pages}{632–635}.
\newblock
\showISBNx{9781450392785}
\urldef\tempurl%
\url{https://doi.org/10.1145/3523227.3551468}
\showDOI{\tempurl}


\bibitem[Lazovich et~al\mbox{.}(2022)]%
        {Lazovich_2022}
\bibfield{author}{\bibinfo{person}{Tomo Lazovich}, \bibinfo{person}{Luca Belli}, \bibinfo{person}{Aaron Gonzales}, \bibinfo{person}{Amanda Bower}, \bibinfo{person}{Uthaipon Tantipongpipat}, \bibinfo{person}{Kristian Lum}, \bibinfo{person}{Ferenc Huszár}, {and} \bibinfo{person}{Rumman Chowdhury}.} \bibinfo{year}{2022}\natexlab{}.
\newblock \showarticletitle{Measuring disparate outcomes of content recommendation algorithms with distributional inequality metrics}.
\newblock \bibinfo{journal}{\emph{Patterns}} \bibinfo{volume}{3}, \bibinfo{number}{8} (\bibinfo{date}{Aug.} \bibinfo{year}{2022}), \bibinfo{pages}{100568}.
\newblock
\showISSN{2666-3899}
\urldef\tempurl%
\url{https://doi.org/10.1016/j.patter.2022.100568}
\showDOI{\tempurl}


\bibitem[Li et~al\mbox{.}(2017)]%
        {10.1145/3123266.3123316}
\bibfield{author}{\bibinfo{person}{Jingjing Li}, \bibinfo{person}{Ke Lu}, \bibinfo{person}{Zi Huang}, {and} \bibinfo{person}{Heng~Tao Shen}.} \bibinfo{year}{2017}\natexlab{}.
\newblock \showarticletitle{Two Birds One Stone: On both Cold-Start and Long-Tail Recommendation}. In \bibinfo{booktitle}{\emph{Proceedings of the 25th ACM International Conference on Multimedia}} (Mountain View, California, USA) \emph{(\bibinfo{series}{MM '17})}. \bibinfo{publisher}{Association for Computing Machinery}, \bibinfo{address}{New York, NY, USA}, \bibinfo{pages}{898–906}.
\newblock
\showISBNx{9781450349062}
\urldef\tempurl%
\url{https://doi.org/10.1145/3123266.3123316}
\showDOI{\tempurl}


\bibitem[Rendle et~al\mbox{.}(2020)]%
        {rendle2020neuralcollaborativefilteringvs}
\bibfield{author}{\bibinfo{person}{Steffen Rendle}, \bibinfo{person}{Walid Krichene}, \bibinfo{person}{Li Zhang}, {and} \bibinfo{person}{John Anderson}.} \bibinfo{year}{2020}\natexlab{}.
\newblock \bibinfo{title}{Neural Collaborative Filtering vs. Matrix Factorization Revisited}.
\newblock
\newblock
\showeprint[arxiv]{2005.09683}~[cs.IR]
\urldef\tempurl%
\url{https://arxiv.org/abs/2005.09683}
\showURL{%
\tempurl}


\bibitem[Rhee et~al\mbox{.}(2022)]%
        {10.1145/3523227.3546757}
\bibfield{author}{\bibinfo{person}{Wondo Rhee}, \bibinfo{person}{Sung~Min Cho}, {and} \bibinfo{person}{Bongwon Suh}.} \bibinfo{year}{2022}\natexlab{}.
\newblock \showarticletitle{Countering Popularity Bias by Regularizing Score Differences}. In \bibinfo{booktitle}{\emph{Proceedings of the 16th ACM Conference on Recommender Systems}} (Seattle, WA, USA) \emph{(\bibinfo{series}{RecSys '22})}. \bibinfo{publisher}{Association for Computing Machinery}, \bibinfo{address}{New York, NY, USA}, \bibinfo{pages}{145–155}.
\newblock
\showISBNx{9781450392785}
\urldef\tempurl%
\url{https://doi.org/10.1145/3523227.3546757}
\showDOI{\tempurl}


\bibitem[Wang et~al\mbox{.}(2021)]%
        {Wang_2021}
\bibfield{author}{\bibinfo{person}{Ruoxi Wang}, \bibinfo{person}{Rakesh Shivanna}, \bibinfo{person}{Derek Cheng}, \bibinfo{person}{Sagar Jain}, \bibinfo{person}{Dong Lin}, \bibinfo{person}{Lichan Hong}, {and} \bibinfo{person}{Ed Chi}.} \bibinfo{year}{2021}\natexlab{}.
\newblock \showarticletitle{DCN V2: Improved Deep \& Cross Network and Practical Lessons for Web-scale Learning to Rank Systems}. In \bibinfo{booktitle}{\emph{Proceedings of the Web Conference 2021}} (Ljubljana, Slovenia) \emph{(\bibinfo{series}{WWW '21})}. \bibinfo{publisher}{Association for Computing Machinery}, \bibinfo{address}{New York, NY, USA}, \bibinfo{pages}{1785–1797}.
\newblock
\showISBNx{9781450383127}
\urldef\tempurl%
\url{https://doi.org/10.1145/3442381.3450078}
\showDOI{\tempurl}


\bibitem[Yang et~al\mbox{.}(2020)]%
        {50257}
\bibfield{author}{\bibinfo{person}{Ji Yang}, \bibinfo{person}{Xinyang Yi}, \bibinfo{person}{Derek~Zhiyuan Cheng}, \bibinfo{person}{Lichan Hong}, \bibinfo{person}{Yang Li}, \bibinfo{person}{Simon Wang}, \bibinfo{person}{Taibai Xu}, {and} \bibinfo{person}{Ed~H. Chi}.} \bibinfo{year}{2020}\natexlab{}.
\newblock \showarticletitle{Mixed Negative Sampling for Learning Two-tower Neural Networks in Recommendations}.
\newblock


\bibitem[Yin et~al\mbox{.}(2012)]%
        {yin2012challenginglongtailrecommendation}
\bibfield{author}{\bibinfo{person}{Hongzhi Yin}, \bibinfo{person}{Bin Cui}, \bibinfo{person}{Jing Li}, \bibinfo{person}{Junjie Yao}, {and} \bibinfo{person}{Chen Chen}.} \bibinfo{year}{2012}\natexlab{}.
\newblock \bibinfo{title}{Challenging the Long Tail Recommendation}.
\newblock
\newblock
\showeprint[arxiv]{1205.6700}~[cs.DB]
\urldef\tempurl%
\url{https://arxiv.org/abs/1205.6700}
\showURL{%
\tempurl}


\end{thebibliography}

\end{document}